# Frustration phenomena in Josephson point contacts between single-band and three-band superconductors

Y.S. Yerin and A.N. Omelyanchouk

*B.Verkin Institute for Low Temperature Physics and Engineering
of the National Academy of Sciences of Ukraine
47 Lenin Ave., 61103 Kharkov, Ukraine*

**Abstract**

Within the formalism of Usadel equations the Josephson effect in dirty point contacts between single-band and three-band superconductors is investigated. The general expression for the Josephson current, which is valid for arbitrary temperatures, is obtained. We calculate current-phase relations for very low temperature and in the vicinity of the critical temperature. For three-band superconductors with broken time-reversal symmetry (BTRS) point contacts undergo frustration phenomena with different current-phase relations, corresponding to φ-contacts. For three-band superconductors without BTRS we have close to sinusoidal current-phase relations and absence of the frustration, excepting the case of very low temperature, where under certain conditions two ground states of the point contact are realized. Our results can be used as the potential probe for the detection of the possible BTRS state in three-band superconducting systems.

## 1. Introduction

The symmetry of the order parameter of recently discovered iron-based superconductors still remains a controversial question and an unresolved challenge. The initial hypothesis that the order parameter has two components with opposite signs ($s_\pm$-wave symmetry) is casted doubt on by numerous data obtained during the ARPES [1] and in experiments on the temperature dependence of the specific heat capacity [2-4]. These results indirectly indicate the presence in these compounds superconducting chiral state like spin-triplet p-wave in strontium ruthenate [5] or recently proposed d + id wave superconductivity in graphene [6].

At the same time it is well known that chiral superconductivity can lead to an interesting phenomenon in such compounds, namely the broken time-reversal symmetry (BTRS): phases of the multicomponent order parameter undergo frustration, leading to the emergence of several "equal in rights" ground states of the superconductor.

BTRS in superconducting oxypnictides and chalcogenides is discussed earlier in the frame of the s+ id [7] and $s_\pm+is_{++}$ [8] symmetry of the order parameter. These models assumed the presence of two components of the order parameter (a two-band superconductor). However, the latest experimental data give clear evidences about the presence of at least three energy gaps in the spectrum of quasiparticle excitations in iron-based superconductors [9, 10]. The presence of three interacting parameters also leads to the BTRS state [11-20].

In this regard, a reasonable question arises about the possible experimental techniques for the creating and subsequent detection of this phenomenon in iron-based superconductors. Currently in this sense the most prominent candidate among known superconducting iron oxypnictides and chalcogenides is $Ba_{1-x}K_xFe_2As_2$, in which BTRS can be achieved by the controlling the level of doping [20]. In turn, methods that have been proposed to reveal this phenomenon use the detection of Legget modes [21], observation of the unusual behavior of the magnetization of the sample in the process of the fast quench cooling [22] and the detection of kinks on the current versus applied magnetic flux dependencies in a doubly-connected mesoscopic sample [23].

We believe that another useful way to detect the state with BTRS in iron-based superconductors is the investigation of the Josephson effect, which is traditionally considered as the most powerful tool for detecting the manifestation of the phase of the order parameter in single and multiband superconductors. Attempt to understand how the frustration of phases of the order parameters effects on current-phase relations of a contact between conventional (s-wave) single-band and three-band superconductor with BTRS one has been undertaken already in [24]. However this investigation was done for the ballistic regime which is difficultly to achieve in real experimental conditions.

In the present paper within the formalism of the Usadel equations [25], generalized for the case of three energy gaps [26], we investigated a dirty point contact between the s-wave single-band superconductor and the three-band one. We found qualitative differences in the structure of the current-phase relations of the point contact for cases of the three-band superconductor with the presence of BTRS and without of this state.

## 2. Ground states of a homogeneous equilibrium three-band superconductor

At the beginning we investigate a homogeneous equilibrium three-band superconductor with strong impurity intraband scattering rates (dirty limit) and without interband scattering in order to find all possible frustrated and non-frustrated ground states. In this limit the three-band superconductor is described by the Usadel equations for normal and anomalous Green's functions $g_i$ and $f_i$:

$$\omega f_i - \frac{1}{2} D_i \left( g_i \nabla^2 f_i - f_i \nabla^2 g_i \right) = \Delta_i g_i, \quad i = 1, 2, 3. \tag{1}$$

Equations (1) must be supplemented with self-consistency equation for order parameters $\Delta_i$:

$$\Delta_i = 2\pi T \sum_j \sum_{\omega>0}^{\langle \omega_0 \rangle} \lambda_{ij} f_j, \tag{2}$$

and the expression for the current density

$$j = -2ie\pi T \sum_i \sum_{\omega>0} N_i D_i \left( f_i^* \nabla f_i - f_i \nabla f_i^* \right). \tag{3}$$

Normal and anomalous Green's functions $g_i$ and $f_i$, which are connected by the normalization condition $g_i^2 + |f_i|^2 = 1$, are functions of coordinates **r** and the Matsubara frequency $\omega = (2n+1)\pi T$. $D_i$ are the intraband diffusivities due to nonmagnetic intraband impurity scattering, $N_i$ are the densities of states on the Fermi surface of the $i$-th band, $\lambda_{ij}$ are BCS interaction constants and $\langle\omega_0\rangle$ is the cut-off frequency.

For the equilibrium homogeneous state Usadel equations (1) have solutions

$$f_i = \frac{|\Delta_i|\exp(i\varphi_i)}{\sqrt{\omega^2 + |\Delta_i|^2}}, \tag{4}$$

where $\varphi_i$ are phases of each order parameter.

Ground states of a three-band superconductor can be found from the minimization of the free energy density in respect to phase differences of the order parameters $\phi = \varphi_1 - \varphi_2$ and $\theta = \varphi_1 - \varphi_3$:

$$F = \frac{1}{2}\sum_{ij}\Delta_i\Delta_j^* N_i \lambda_{ij}^{-1} + \sum_i F_i, \tag{5}$$

where $\lambda_{ij}^{-1}$ is the inverse matrix of interaction constants $\lambda_{ij}$ and

$$F_i = 2\pi T \sum_{\omega>0}^{\langle\omega_0\rangle} N_i \left[\omega(1-g_i) - \text{Re}(f_i^*\Delta_i) + \frac{1}{4}D_i(\nabla f_i \nabla f_i^* + \nabla g_i \nabla g_i)\right] \tag{6}$$

represents intraband energies of the three-band superconductor.

For $i = 1, 2$ we obtain the free energy density of a two-band superconductor, which was used for the prediction of phase textures in multi-band and multi-band-like superconducting systems [27]. The first term in (5) contains three interband (Josephson-like) interaction energies $\gamma_{12}\cos\phi$, $\gamma_{13}\cos\theta$ and $\gamma_{23}\cos(\theta-\phi)$ where $\gamma_{ij} = -\lambda_{ji}^{-1}N_j$ (usually $\lambda_{ij}^{-1}N_i = \lambda_{ji}^{-1}N_j$, $i \neq j$) are interband interaction coefficients, which are used in Ginzburg-Landau approach; for $\gamma_{ij} > 0$ attractive interband interactions are took place, while for $\gamma_{ij} < 0$ interactions are repulsive.

The first variation of (5) on $\phi$ and $\theta$ gives:

$$-(\lambda_{12}^{-1}N_1 + \lambda_{21}^{-1}N_2)|\Delta_1||\Delta_2|\sin\phi + (\lambda_{23}^{-1}N_2 + \lambda_{32}^{-1}N_3)|\Delta_2||\Delta_3|\sin(\theta-\phi) = 0, \tag{7}$$

$$-(\lambda_{13}^{-1}N_1 + \lambda_{31}^{-1}N_3)|\Delta_1||\Delta_3|\sin\theta - (\lambda_{23}^{-1}N_2 + \lambda_{32}^{-1}N_3)|\Delta_2||\Delta_3|\sin(\theta-\phi) = 0. \tag{8}$$

Solutions of (7) and (8) for $\phi$ and $\theta$, which determine the points of extremum, depend from their arrangement in quadrants.

Introducing $\Omega = \sqrt{1 - \left(\dfrac{G_3^2 G_2^2 |\Delta_3|^2 - G_1^2 G_3^2 |\Delta_1|^2 - G_1^2 G_2^2 |\Delta_2|^2}{2G_1^2 G_2 G_3 |\Delta_1||\Delta_2|}\right)^2}$, where $G_1 = \lambda_{12}^{-1} N_1 + \lambda_{21}^{-1} N_2$,

$G_2 = \lambda_{23}^{-1} N_2 + \lambda_{32}^{-1} N_3$ and $G_3 = \lambda_{13}^{-1} N_1 + \lambda_{31}^{-1} N_3$ for $\phi \in \left[-\dfrac{\pi}{2}, \dfrac{\pi}{2}\right]$ and $\theta \in \left[-\dfrac{\pi}{2}, \dfrac{\pi}{2}\right]$ we have

$$\begin{cases} \phi = \pm \arcsin \Omega, \\ \theta = \mp \arcsin\left(\dfrac{G_1 |\Delta_2|}{G_3 |\Delta_3|} \Omega\right), \end{cases} \quad \begin{cases} \phi = 0, \\ \theta = 0, \end{cases} \tag{9}$$

for $\phi \in \left[\dfrac{\pi}{2}, \dfrac{3\pi}{2}\right]$ and $\theta \in \left[-\dfrac{\pi}{2}, \dfrac{\pi}{2}\right]$

$$\begin{cases} \phi = \pi \pm \arcsin \Omega, \\ \theta = \pm \arcsin\left(\dfrac{G_1 |\Delta_2|}{G_3 |\Delta_3|} \Omega\right), \end{cases} \quad \begin{cases} \phi = \pi, \\ \theta = 0, \end{cases} \tag{10}$$

for $\phi \in \left[-\dfrac{\pi}{2}, \dfrac{\pi}{2}\right]$ and $\theta \in \left[\dfrac{\pi}{2}, \dfrac{3\pi}{2}\right]$

$$\begin{cases} \phi = \pm \arcsin \Omega, \\ \theta = \pi \pm \arcsin\left(\dfrac{G_1 |\Delta_2|}{G_3 |\Delta_3|} \Omega\right), \end{cases} \quad \begin{cases} \phi = 0, \\ \theta = \pi, \end{cases} \tag{11}$$

and for $\phi \in \left[\dfrac{\pi}{2}, \dfrac{3\pi}{2}\right]$ and $\theta \in \left[\dfrac{\pi}{2}, \dfrac{3\pi}{2}\right]$

$$\begin{cases} \phi = \pi \pm \arcsin \Omega, \\ \theta = \pi \mp \arcsin\left(\dfrac{G_1 |\Delta_2|}{G_3 |\Delta_3|} \Omega\right), \end{cases} \quad \begin{cases} \phi = \pi, \\ \theta = \pi. \end{cases} \tag{12}$$

For given $\lambda_{ij}$ and computed for these values $|\Delta_i|$ by Eqs. (2) and (4) we have eight possible solutions for $\phi$ and $\theta$ (9-12). Selection of the proper solution, which corresponds to the ground state, is provided by the condition for the minimum of $F(\theta, \phi)$, following from the second variation of the free energy density (5). Final form of the expression for the second variation also depends on the arrangement in quadrants of $\phi$ and $\theta$.

### 3. Josephson current between single-band and three-band superconductors

The point contact can be considered as a weak superconducting link in the form of thin filament of the length $L$ and diameter $d$, connecting two superconducting bulk banks (fig.1).

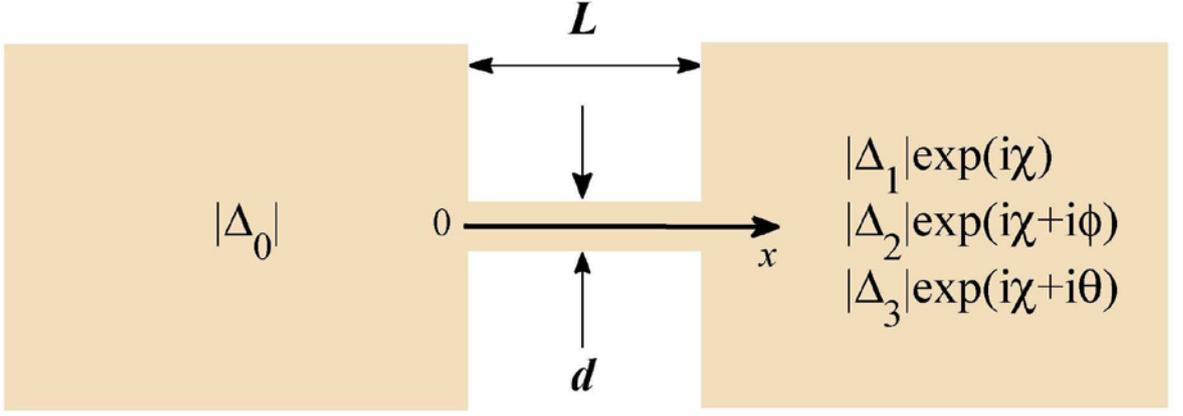

Fig. 1. The model of the point contact between bulk single-band and three-band superconductors as two banks connected by the thing filament of a length $L$ and a diameter $d$.

On conditions that $d \ll L$ and $d \ll \min \xi_i(T)$ ($\xi_i(T)$ - coherence lengths in the $i$-th band) we can solve a one-dimensional problem inside the filament ($0 \leq x \leq L$) and neglect all terms in Usadel equations (1) except the gradients ones. Using the normalization condition we have equations for $f_i$

$$\sqrt{1-|f_i|^2}\frac{d^2}{dx^2}f_i - f_i\frac{d^2}{dx^2}\sqrt{1-|f_i|^2} = 0, \ i=1,2,3. \tag{13}$$

The boundary conditions for equations (13) at $x = 0, L$ are determined by the values of $f_i$ in banks:

$$f_i(0) = \frac{|\Delta_0|}{\sqrt{|\Delta_0|^2 + \omega^2}}, \tag{14}$$

$$f_1(L) = \frac{|\Delta_1|\exp(i\chi)}{\sqrt{|\Delta_1|^2 + \omega^2}}, \ f_2(L) = \frac{|\Delta_2|\exp(i\chi+i\phi)}{\sqrt{|\Delta_2|^2 + \omega^2}}, \ f_3(L) = \frac{|\Delta_3|\exp(i\chi+i\theta)}{\sqrt{|\Delta_3|^2 + \omega^2}}, \tag{15}$$

where $\chi$ is the phase difference between the first order parameter of the three-band superconductor and the order parameter of the single-band one and where $\phi$ and $\theta$ determine phase differences (ground state) in the bulk three-band superconductor.

Eqs. (13) admit analytical solution with boundary conditions (14), (15). Taking into account expression for the current density (3) we get for the Josephson current between the single-band and the three-band superconductor

$$I = \sum_i I_i, \tag{16}$$

where

$$I_i = \frac{2\pi T}{eR_{Ni}}b_i\sum_{\omega>0}\frac{1}{p_i}\left[\arctan\left(\frac{\Delta_0 d_i - a_i b_i}{p_i}\right) + \arctan\left(\frac{\Delta_i d_i + a_i b_i}{p_i}\right)\right]. \tag{17}$$

Here $R_{Ni}$ are partial contributions to the point contact resistance. Also notations

$$p_i = \sqrt{b_i^2 + (a_i^2+1)\omega^2}, a_1 = \frac{|\Delta_1|-|\Delta_0|}{|\Delta_1|+|\Delta_0|}\cot\frac{\chi}{2}, \quad a_2 = \frac{|\Delta_2|-|\Delta_0|}{|\Delta_2|+|\Delta_0|}\cot\frac{\chi+\phi}{2}, \quad a_3 = \frac{|\Delta_3|-|\Delta_0|}{|\Delta_3|+|\Delta_0|}\cot\frac{\chi+\theta}{2},$$

$$b_1 = \frac{2|\Delta_0||\Delta_1|}{|\Delta_0|+|\Delta_1|}\cos\frac{\chi}{2}, \quad b_2 = \frac{2|\Delta_0||\Delta_2|}{|\Delta_0|+|\Delta_2|}\cos\frac{\chi+\phi}{2}, \quad b_3 = \frac{2|\Delta_0||\Delta_3|}{|\Delta_0|+|\Delta_3|}\cos\frac{\chi+\theta}{2} \quad d_1 = (a_1^2+1)\sin\frac{\chi}{2},$$

$d_2 = (a_2^2+1)\sin\frac{\chi+\phi}{2}$ and $d_3 = (a_3^2+1)\sin\frac{\chi+\theta}{2}$ were used.

For $i=1$ we get expression for the Josephson current between single-band superconductors [28, 29], which for coinciding values of energy gaps turns into Kulik-Omelyanchouk theory for dirty point contacts [30].

### 3.1. Josephson current for $T = 0$

For $T=0$ in the expression (17) we can turn from the summation over Matsubara frequencies to the integration and get for the total current

$$I = \frac{\pi}{eR_{N2}}\frac{|\Delta_0||\Delta_1|}{|\Delta_0|+|\Delta_1|}\frac{\cos\frac{\chi}{2}}{\sqrt{a_1^2+1}}\ln Q_2 + \frac{\pi}{eR_{N2}}\frac{|\Delta_0||\Delta_2|}{|\Delta_0|+|\Delta_2|}\frac{\cos\left(\frac{\chi+\phi}{2}\right)}{\sqrt{a_2^2+1}}\ln Q_2$$
$$+ \frac{\pi}{eR_{N3}}\frac{|\Delta_0||\Delta_3|}{|\Delta_0|+|\Delta_3|}\frac{\cos\left(\frac{\chi+\theta}{2}\right)}{\sqrt{a_3^2+1}}\ln Q_3. \quad (18)$$

Here

$$Q_i = \frac{\left[|\Delta_0|d_i - a_ib_i + \sqrt{(|\Delta_0|d_i - a_ib_i)^2 + b_i^2}\right]\left[|\Delta_i|d_i + a_ib_i + \sqrt{(|\Delta_i|d_i + a_ib_i)^2 + b_i^2}\right]}{b_i^2}. \quad (19)$$

In the following we consider for the simplicity the case of coinciding energy gaps $|\Delta_0|=|\Delta_1|=|\Delta_2|=|\Delta_3|=|\Delta|$ and obtain for the total current

$$I = \frac{\pi|\Delta|}{eR_{N1}}\cos\frac{\chi}{2}\operatorname{arctanh}\sin\frac{\chi}{2} + \frac{\pi|\Delta|}{eR_{N2}}\cos\frac{\chi+\phi}{2}\operatorname{arctanh}\sin\frac{\chi+\phi}{2}$$
$$+ \frac{\pi|\Delta|}{eR_{N3}}\cos\frac{\chi+\theta}{2}\operatorname{arctanh}\sin\frac{\chi+\theta}{2}. \quad (20)$$

Integrating (20) over $\chi$ we obtain the expression for the Josephson energy of the point contact

$$E = \frac{|\Delta|\Phi_0}{2eR_{N1}}\left(2\sin\frac{\chi}{2}\operatorname{arctanh}\sin\frac{\chi}{2} + \ln\cos^2\frac{\chi}{2}\right) + \frac{|\Delta|\Phi_0}{2eR_{N2}}\left(2\sin\frac{\chi+\phi}{2}\operatorname{arctanh}\sin\frac{\chi+\phi}{2} + \ln\cos^2\frac{\chi+\phi}{2}\right)$$
$$+ \frac{|\Delta|\Phi_0}{2eR_{N3}}\left(2\sin\frac{\chi+\theta}{2}\operatorname{arctanh}\sin\frac{\chi+\theta}{2} + \ln\cos^2\frac{\chi+\theta}{2}\right).$$
$$(21)$$

We can select arbitrary values of $\phi$ and $\theta$ because it's possible to match appropriate values of $\lambda_{ij}$ to satisfy the self-consistency equation (3) and expressions (9)-(12). In other words, we have only five equations, three of which follow from the self-consistency equation (3) and two from expressions for $\phi$, $\theta$, for the determination of nine variables $\lambda_{ij}$. Based on these arguments we assume that for the frustrated three-band superconductor one of the ground state can be, for instance, $\phi = 0.6\pi$, $\theta = 1.2\pi$. Since these phase differences were chosen in the second and in the third quadrants respectively according to the solution (12) another ground state should correspond to such values: $\phi = 1.4\pi$, $\theta = 0.8\pi$.

So when one of contacting banks is the three-band superconductor with BTRS state we observe complicated current-phase relations with the behavior of Josephson energies, corresponding to $\varphi$-contact (fig. 2), following to the terminology after [31]. Here and hereinafter we will call $\varphi$-contact as a type of the Josephson junction with an arbitrary phase shift $\varphi$ in the ground state.

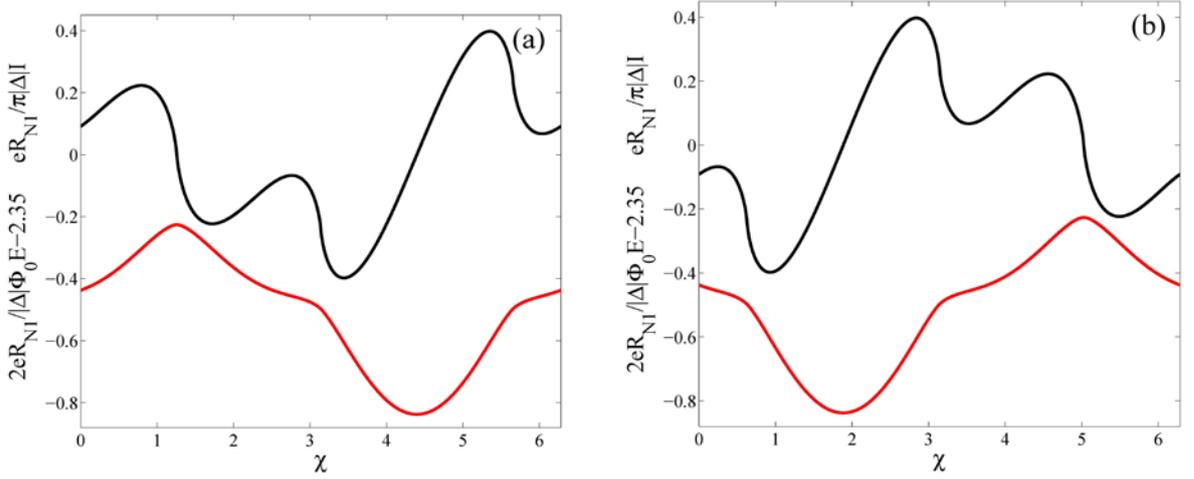

Fig. 2. Current-phase relations (black lines) and Josephson energies (red lines) of point contacts between single-band and three-band superconductor with BTRS in the case of coinciding energy gaps for $\phi = 0.6\pi$, $\theta = 1.2\pi$ (**a**) and $\phi = 1.4\pi$, $\theta = 0.8\pi$ (**b**). Ratios $R_{N1}/R_{N2} = R_{N1}/R_{N3} = 1$.

So during experimental measurements for the same BTRS three-band superconductor we can observe different current-phase relations. It depends from the "prehistory" of the three-band superconductor, i.e. how was the ground state for this superconducting system achieved.

Also we consider the simplest case, which is often cited for the illustration of BTRS in three-band superconductors, when such systems have the odd number of repulsive interband interactions $\gamma_{ij}$ with equal modules. In this case two ground states are possible: $(\phi, \theta) = (-\pi/3, \pi/3)$ and $(\phi, \theta) = (\pi/3, -\pi/3)$, if we have only one repulsive interband interaction and two attractive ones and $(\phi, \theta) = (-2\pi/3, 2\pi/3)$ and $(\phi, \theta) = (2\pi/3, -2\pi/3)$, if all interband interactions are repulsive.

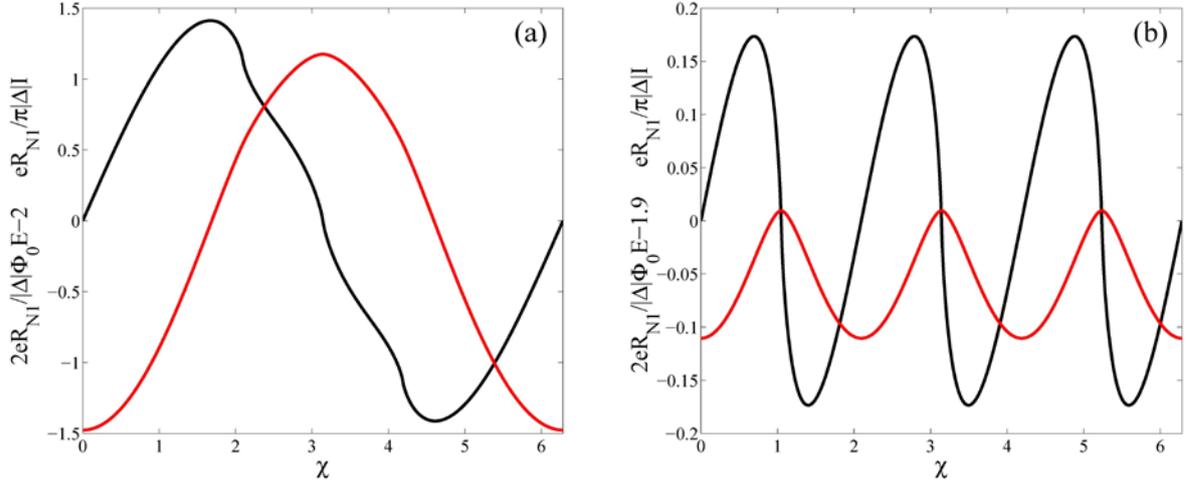

Fig. 3. The same as in fig. 2 for $\phi = \pi/3$, $\theta = -\pi/3$; $\phi = -\pi/3$, $\theta = \pi/3$ (**a**) and $\phi = 2\pi/3$, $\theta = -2\pi/3$; $\phi = -2\pi/3$, $\theta = 2\pi/3$ (**b**). Ratios $R_{N1}/R_{N2} = R_{N1}/R_{N3} = 1$.

Firstly we found that for these two ground states current-phase relations and Josephson energies coincide (see fig. 3) in comparison with above considered case (fig. 2). This fact can be easily understood from expressions (20) and (21) bearing in mind that the cosine is even function and the inverse hyperbolic tangent and the sine are odd ones. Secondly, despite the presence of the BTRS state in the three-band superconductor the most remarkable feature for $(\phi,\theta) = (-\pi/3, \pi/3)$ and $(\phi,\theta) = (\pi/3, -\pi/3)$ is that there is no frustration of the point contact (fig. 3a) with inflection points in the middle of the current-phase relation curve. Thirdly, for $(\phi,\theta) = (-2\pi/3, 2\pi/3)$ and $(\phi,\theta) = (2\pi/3, -2\pi/3)$ we have current-phase relation with triply degenerates states (fig. 3b), i.e. frustration with three ground states of the point contact is occurred.

The current-phase relation (20) and the Josephson energy (21) for the point contact between single-band superconductor and non-BTRS three-band one are shown on the fig. 4. Here the peculiarity of such point contacts are the occurrence of a $\varphi$-contact with frustration of ground states (fig. 4b and 4c) as for the BTRS case if the three-band superconductor has the ground state for $\phi = 0$, $\theta = \pi$ or $\phi = \theta = \pi$.

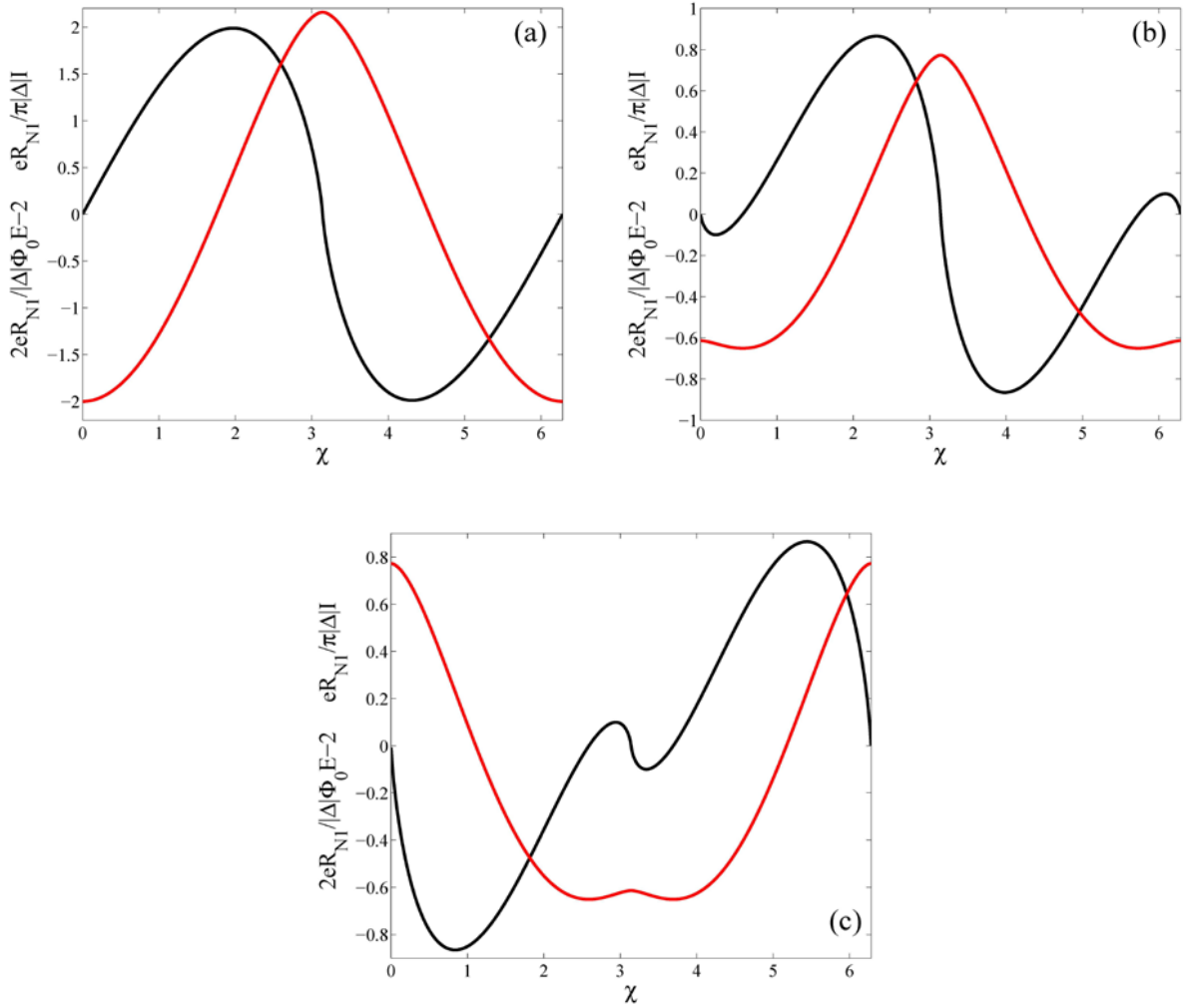

Fig. 4. Current-phase relations (black lines) and Josephson energies (red lines) of point contacts between single-band and three-band superconductor without BTRS in the case of coinciding energy gaps for $\phi = \theta = 0$ (**a**) and $\phi = 0$, $\theta = \pi$ (**b**, for $\phi = \pi$ and $\theta = 0$ it will be the same dependence) and $\phi = \theta = \pi$ (**c**). Ratios $R_{N1}/R_{N2} = R_{N1}/R_{N3} = 1$.

Thus at $T = 0$ in both cases of BTRS and non-BTRS three-band superconductors we can have frustration phenomenon in point contacts. In order to distinguish three-band superconductors with and without BTRS state in the next section we consider point contacts in the vicinity of the critical temperature $T_c$ (it can be the critical temperature of the single- or the three-band superconductor in dependence on what is the value lower).

### 3.2. Josephson current in the vicinity of the critical temperature

By the linearization of the expression (17) we get for total current

$$I = \frac{\pi |\Delta|^2}{4eT_c R_{N1}} \sin \chi + \frac{\pi |\Delta|^2}{4eT_c R_{N2}} \sin(\chi + \phi) + \frac{\pi |\Delta|^2}{4eT_c R_{N3}} \sin(\chi + \theta), \qquad (22)$$

and after integrating over $\chi$ for the Josephson energy

$$E = -\frac{|\Delta|^2 \Phi_0}{8eT_c R_{N1}} \cos \chi - \frac{|\Delta|^2 \Phi_0}{8eT_c R_{N2}} \cos(\chi + \phi) - \frac{|\Delta|^2 \Phi_0}{8eT_c R_{N3}} \cos(\chi + \theta). \tag{23}$$

For the three-band superconductor with BTRS the intricate behavior of $I(\varphi)$ and $E(\varphi)$ dependencies (fig. 2) turns into simple sinusoidal forms but nevertheless with the conservation of a $\varphi$-contact feature (fig. 5).

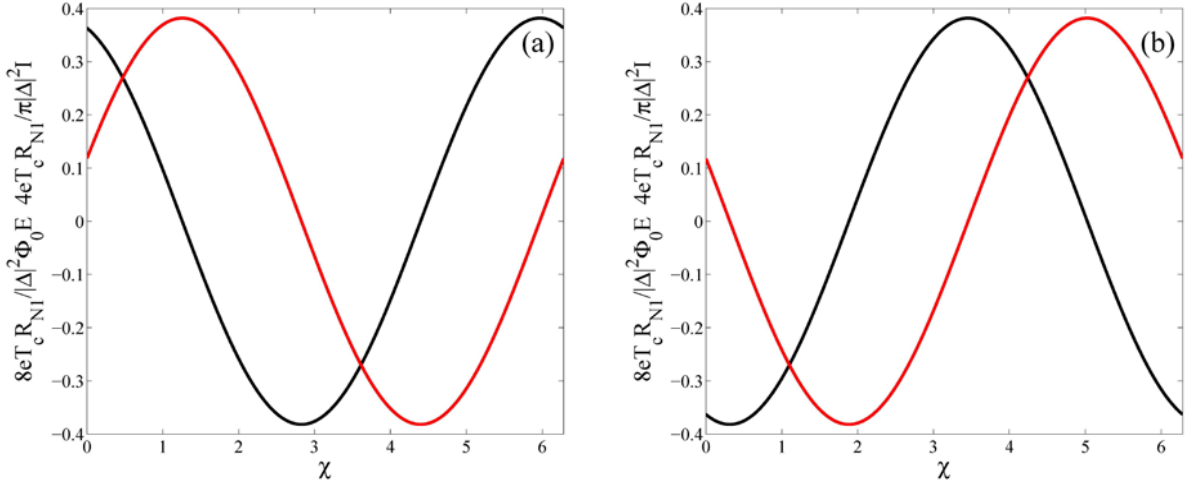

Fig. 5. Current-phase relations (black lines) and Josephson energies (red lines) of point contacts between single-band and three-band superconductor with BTRS in the case of coinciding energy gaps and in the vicinity of the critical temperature for $\phi = 0.6\pi$, $\theta = 1.2\pi$ (a) and $\phi = 1.4\pi$, $\theta = 0.8\pi$ (b). Ratios $R_{N1}/R_{N2} = R_{N1}/R_{N3} = 1$.

For the point contact when one of the bank is the three-band superconductor with one repulsive interband interaction and with equal modules of $\gamma_{ij}$ current-phase relations continue to coincide, but now lost inflection points, which are took place for the very low temperature (fig. 3a) and transform to clear sinusoidal dependence (fig. 6). At the same time the ground state is not varied and the point contact remains conventional.

If one of the contacting banks is the three-band superconductor with all repulsive interband interaction and again with equal modules of $\gamma_{ij}$ the point contact is characterized by the zero Josephson current for both ground states of such three-band superconductor. It seems something unexpectedly but if we substitute $\phi = -2\pi/3$, $\theta = 2\pi/3$ (for $\phi = 2\pi/3$, $\theta = -2\pi/3$ it would be the same) into the expression for the current (22) after the simplification we get zero current.

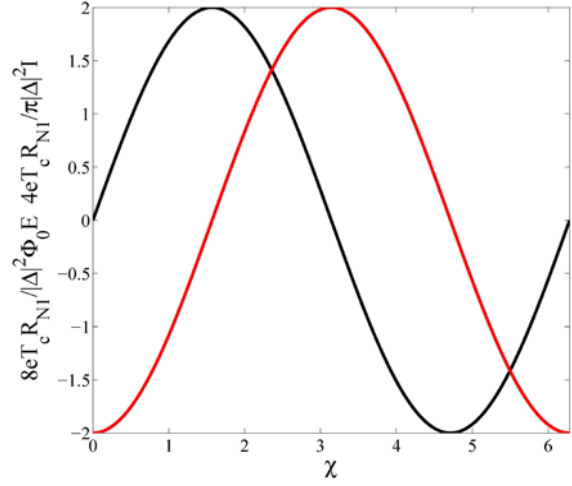

Fig. 6. The same as in fig. 5 for $\phi = \pi/3$, $\theta = -\pi/3$ and $\phi = -\pi/3$, $\theta = \pi/3$. For $\phi = 2\pi/3$, $\theta = -2\pi/3$ and $\phi = -2\pi/3$, $\theta = 2\pi/3$ absence of the Josephson current is took place (see explanation in the text).

Current-phase relations (22) and Josephson energies (23) in the case of the three-band superconductor without BTRS are shown on the fig. 7.

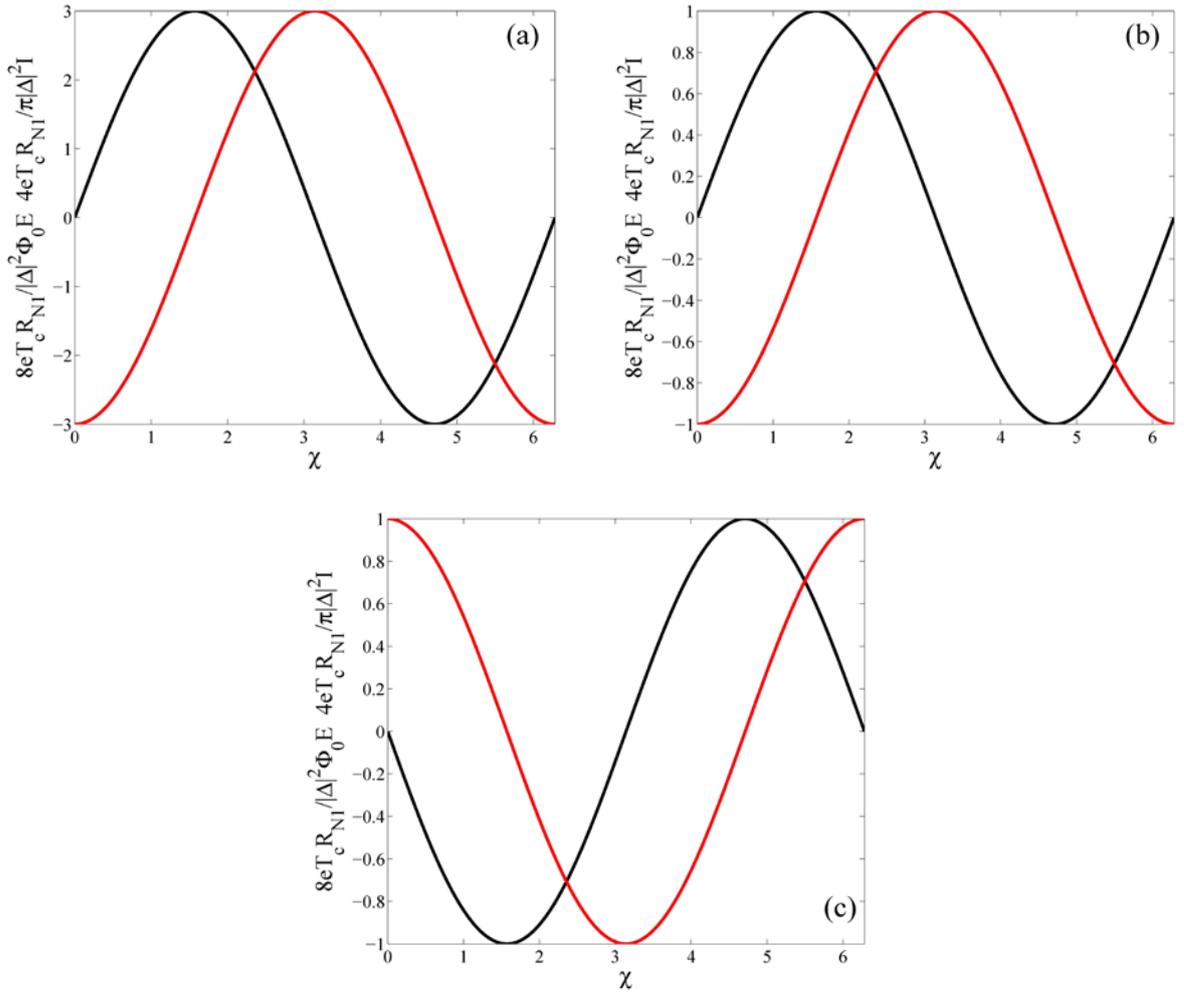

Fig.7. Current-phase relations (black lines) and Josephson energies (red lines) of point contacts between single-band and three-band superconductors without BTRS in the case of coinciding energy gaps and in the vicinity of the critical temperature for $\phi = \theta = 0$ (**a**) and $\phi = 0$, $\theta = \pi$ (**b**, for $\phi = \pi$ and $\theta = 0$ it will be the same dependence) and $\phi = \theta = \pi$ (**c**). Ratios $R_{N1}/R_{N2} = R_{N1}/R_{N3} = 1$.

Figure 7 demonstrates that for the non-frustrated three-band superconductor with $\phi = 0$, $\theta = \pi$ and $\phi = \theta = \pi$ the proximity of the critical temperature removes a degeneracy of the ground state of the point contact transforming $\varphi$-contact to conventional one (if $\phi = 0$, $\theta = \pi$) or to π-contact (if $\phi = \theta = \pi$).

Comparing current-phase relations of point contacts we can definitely claim the difference between three-band superconductors with the BTRS state and without one. From the experimental point of view the identification procedure can be done in the following way: if a point contact demonstrates two different current-phase relations with the properties of a φ-contact at very low temperature and in the vicinity of $T_c$ during several repeating measurements, unambiguously this three-band superconductor has the state with BTRS. Otherwise even if we observe a φ-contact at the temperature close to the zero but conventional or π-contact near $T_c$ a three-band superconductor has no BTRS state.

For the special case of a three-band superconductor with odd number of repulsive interband interactions and equal modules of $\gamma_{ij}$ the detection procedure undergoes changes. During measurements we will observe conventional current-phase relations with inflection points at very low temperature, which disappear near $T_c$ if a three-band superconductor has one repulsive interband interaction and two attractive, or we will observe current-phase relations with triply degenerate ground states at $T = 0$ and zero Josephson current in the vicinity of the critical temperature if all interband interactions are repulsive.

## Conclusions

Based on the microscopic approach, we have obtained general analytical expressions for phase differences of order parameters, corresponding to the ground state of a homogeneous equilibrium three-band superconductor. We have developed microscopic theory of the Josephson effect in dirty point contacts between single-band and three-band superconductors. For a BTRS three-band superconductor we have revealed frustration phenomenon of the point contact with different current-phase relations. By analyzing the Josephson energy we have found that the contact has shown the property of a φ-contact for whole temperature interval from zero to the critical temperature. For a three-band superconductor, which is characterized by the absence of the BTRS state, with the increasing of the temperature we have observed the evolution of the contact behavior from the frustrated φ-contact to conventional or π-contact in dependence on the values of phase differences in a three-band superconductor. We stress that our theoretical results can be useful in experiments on the detection of BTRS states in multi-band superconductors.

## Acknowledgments

This work was supported by DKNII (Project No. M/231-2013).